\documentstyle{lamuphys}
\makeatletter
\let\chapter\hid@chapter
\makeatother
\def\kms{~km s$^{-1}$~}
\def\uas{~$\mu$as~}
\begin{document}
\pagenumbering{arabic}
\titlerunning{Cosmology with the VLTI: Unleashing the Giant}
\title{Cosmology with the Very Large Telescope Interferometer
using a space based astrometric reference frame
\footnote{
Talk at ESO Workshop April 1st 1996,
to appear in: {\sl Science with the VLTI}
(1997) ed. F. Paresce, (Springer: Heidelberg) 
}}

\author{David\,Tytler\inst{}}

\institute{University of California San Diego, CASS 0424, La Jolla
CA 92093-0424, USA}

\maketitle

\begin{abstract}
Cosmology with large interferometric telescopes
is a rich and largely unexplored subject, involving three types of
measurement: astrometric measurement of absolute distances and proper motions, 
dispersions of relative proper motions, and images.
The ground based interferometers can have huge apertures, which are necessary 
for faint cosmological targets. But, alone, they 
are limited to astrometry within the isoplanatic patch, and hence to relative 
positions, which are of little use for parallaxes and proper motions because 
reference stars have unknown parallaxes and
huge (500 $\mu$arcsec) unknown motions.
We propose that space missions should measure global astrometric
parallaxes and proper motions for (V$>16$) reference stars within 
the isoplanatic patches of important cosmological and Galactic targets. 
Ground based interferometers can then
measure absolute distances (parallaxes) and proper motions to 10 $\mu$arcsec,
tied to these reference stars.
In combination, space and ground based interferometers can make a
wide variety of measurements, some of which were believed to be restricted to
space missions, and others where not considered possible because space 
missions lack light gathering power:
absolute distances accurate to $<$10\% for most globular clusters and 
about ten near by galaxies; 
proper motions of stars in near by dwarf galaxies
and stars near to giant black holes;
the masses and distances to individual MACHOS which cause microlensing 
events in the halo and bulge of our galaxy; 
proper motions of 1000 galaxies out to Virgo;
and images of giant black holes, AGN and distant galaxies.

But cosmological observations stretch the VLTI technically.
To observe the few best targets, we need to be able to measure positions to 
$<10 \mu$arcsec over a large portion of the sky.
Since natural guide stars are too far apart, or too faint,
laser guide stars are needed to correct the wavefronts of the
individual 8-m unit telescopes, and the fringe tracking
system must have an extremely high throughput to work on the brightest
stars (V $>16$) near to important targets.
Most of the science is at 1 --2 microns, where excellent 
adaptive optics will be needed on the 8-m telescopes.

\end{abstract}
\section{VLTI Characteristics}
I preview cosmological observations which could be made with ground based 
interferometers which use large aperture telescopes, with specific reference to
the VLTI. I have borrowed heavily from Peterson et al (1996) who reviewed 
science with a space interferometer, because I show that the ground based 
telescopes can undertake some projects which were considered for space alone.
Science with interferometric images with 1 -- 10 $\mu$-arcsecond (\uas)
resolution was discussed by Begelman \& Krolik (1991).
I take an optimistic view, skip over the many
technical and observational difficulties, such as integration times
and confusion in complex fields, and omit many relevant references.
Let us begin with an introduction to the technical issues, which 
require close attention if the VLTI is to do cosmology.

Following the Palomar Testbed Interferometer (\cite{sc}),
the VLTI will have a dual feed, which allows high precision narrow
angle (or differential) astrometry (Kovalevsky 1995). 
One feed looks at a relatively bright reference
star, while the other looks at the target, which can be very faint.
The two can be separated by up to 60 arcsec, but astrometric 
accuracy improves linearly with the inverse separation (\cite{sc}), 
as the paths through the atmosphere to the target and reference 
become nearly identical.
Each feed covers about 2 arcsec diameter, the potential imaging area. 

To correct the wavefront of the light gathered by the individual 8-m 
unit telescopes, we need a bright star V,K $<13$ within 10 arcsec
of the target to sense the wavefront. This restriction rules out all but 
a tiny fraction of the sky, and must be overcome using laser guide stars, 
one for each unit telescope. Lasers allow us to correct the wavefront 
anywhere in the sky.

To provide stable interference fringes (fringe tracking), we can use 
slightly fainter reference 
stars (V,K $<16$ assuming the wavefront is already 
corrected to near the diffraction limit of the unit telescopes). 
Sky coverage as a function of angle to 
the reference star is reasonable at the Galactic plane:
15\% at 10 arcsec, 45\% at 20 arcsec and 75\% at 30 arcsec, but 
unacceptably small at the pole:
0.8\% at 10 arcsec, 3\% at 20 arcsec and 7\% at 30 arcsec.
The ideal fringe tracking system would work on fainter stars,
with higher throughput, wider bandwidth or perhaps better open loop stability.

The key advantage of the VLTI over most other interferometers
is its sensitivity: $V=25$ and $K=23$ in 600 sec with signal to noise
ratio 10 in 0.7 arcsec seeing. This is why it can play an 
important role in cosmology.

Tip/tilt correction 
is sufficient to make the auxiliary 1.8~m telescopes diffraction 
limited at $>1 $ micron, and the unit 8-m telescopes at $>$ 5 microns.
But adaptive optics is essential for cosmology, which will use the
8-m unit telescopes (UTs) at 1 --2 microns.

Baselines are $B<200$ m for the 1.8~m telescopes (0.001 arcsecond = 1 mas
point spread function at 1 micron)
and $B<130$ m for the UTs ($<$1.6 mas). 

\section{Ground Based Astrometry using stars on a Global Reference 
Frame from Space}

In an extremely important paper which opened up a whole new astronomical
technique, Shao \& Colavita (1992)
pointed out that ground based telescopes can make extremely high
precision relative astrometric measurements 
whenever references stars are a few arcsec from targets:
10 \uas with the VLTI for reference stars within $<10$ arcsec, when the
signal to noise in the fringe visibility is about $>$55 at 2.2 microns.
The smaller then angle to the reference, the more similar the light
paths through the atmosphere, and the better the relative
astrometric accuracy.
There are many uses for relative astrometry: binary stars, orbits of
unseen brown dwarfs and planets, masses and distances to microlenses, 
dispersions of the proper motions of stars in clusters and galaxies.

However distances require absolute parallaxes. Parallaxes relative to a
reference star with an unknown parallax are of little use because we do not
know how to proportion the motion between the two stars, and if both are
at the same distance, no relative parallax motion is seen.
Similarly, absolute proper motions are much more useful than those measured
relative to stars with unkown motions.  All faint reference stars move, by about
500 \uas/yr (24 km/s at 10 kpc, which is much more than the 150 \uas 
relative astrometric accuracy of a single 8-m aperture).
There are far too few within an isoplanatic patch to give a reference frame 
stable to 10 \uas, so ground based observations give relative not absolute 
parallaxes and proper motions.

We propose that ground based telescopes should
use reference stars which are observed from space. Space observations 
can give even higher astrometric accuracy over both narrow and wide angles, 
and hence they can construct a global astrometric reference frame, with
absolute parallaxes and motions.
This synergy between the global reference frame from space, and the
fainter targets from the ground based, allows large ground based telescopes 
to measure distances (absolute parallaxes) and absolute proper motions, tied
to the space frame. This greatly increases the range of cosmological
observations for the VLTI.

Of the various space astrometry missions, Hipparchos lacks accuracy: 
1000 \uas.  The proposed ESA mission GAIA is close to what is needed: 
10 \uas to V =15, but it covers the whole sky, where as the VLTI would prefer
fainter reference stars ($V=16$ or 17, as faint as fringe tracking allows), 
selected because they are in the isoplanatic patches of important Galactic and
cosmological
targets. This is a task better suited to NASA's SIM mission, which 
should obtain 5 \uas on pre-selected targets (V=16 in 2,000 sec,
or V=18 in 10,000 seconds, with a baseline of B=7~m).

The use of a space based absolute reference frame is a natural
way to use the VLTI, since there  must be a relatively bright reference
star within the isoplanatic patch for fringe tracking. We are
proposing that this same star should also provide absolute
position, motion and distance.

\section{Comparison with other Instruments}

The VLTI has unique advantages, but only if it is fully equipped with
laser guide stars, adaptive optics, and high throughput fringe
tracking, and even then only for a few 
years, as many other instrument and missions are competitive. 

HIPPARCHOS: 1000 \uas (VLTI 100 times better), survey of $10^5$ stars
(VLTI gives 100 times better relative astrometry, but points at
selected targets).

HST: point spread function of 100 mas at 1 micron 
(VLTI 50 times better), arcmin field of view (VLTI 1000 times smaller).

DIVA: A proposed small German interferometric satellite, which would survey 
$10^6$ stars with V $< 10$ giving parallaxes to 800 \uas (\cite{bas}).

SIM: NASA's Space Interferometer Mission (2003) (\cite{sim}).
V$<20$ (VLTI readily goes 5 magnitudes fainter, but SIM may reach V $=26$ in 20 
hours), 5 yr life (for proper motion -- not parallax -- the VLTI can 
compensate for its lower annual accuracy with more years of observation)
0.3 arcsec field (VLTI is 40 times larger), 4 \uas with a B=10 m
(VLTI is 2 times worse, and needs a space reference grid for parallax),
good UV plane coverage (VLTI is not as good for complex fields),
dynamic range $<2000$ and can null out bright sources to 10$^4$ at 
70 \uas or $10^6$ at 7 \uas (VLTI worse).

GAIA: ESO proposed sky survey to V $< $16 (VLTI 9 magnitudes better), 
$5 \times 10^7$ star sky survey (VLTI points), 
10 \uas at V=15 (VLTI similar, but needs space
reference frame for parallax).

NGST: An 8-m version of NASA's Next Generation Space Telescope 
(www:// ngst.gsfc.nasa.gov/) gives 5 \uas astrometry (VLTI 2 times worse), over
4 arcmin field (VLTI $10^4$ times smaller area),
with much higher sensitivity and dynamic range.
The NGST is dramatic! It will compete strongly with all large
ground based optical/IR telescopes and planned interferometers.

Keck Interferometer: Two 10-m telescopes and four 2-m outriggers, 2003.
(VLTI much better UV coverage).

There are also many interferometers
with smaller apertures including the Palomar Testbed Interferometer,
I2I, ISI, OCAST, SUSI, IOTA, NPOI and CHARA, which will not compete
on fainter cosmology targets.

\section{Assumptions}
For the remainder we will assume a cosmology friendly VLTI.
1) Laser guide stars on at least two 8-m UTs.
2) Adaptive optics on at least two UTs (movable to others), allowing fringe 
tracking on V$>16$ to get good sky coverage, and science at 1 -- 2 microns.
3) Parallax and proper motion from space and on a global reference frame
for at least one reference star (V $\simeq 16$)
in the isoplanatic patch of all interesting targets.
The VLTI then gives positions to about 14 \uas when the reference star is
10 arcsec away, and 30 \uas at 30 arcsec.

\section{Extragalactic Distance Scale} 

The VLTI can be used to provide improved distances to standards.
However space missions such as
GAIA or SIM  are needed to provide reference stars in the global reference 
frame, and those same missions can and will also observe the luminosity 
standards themselves, because they are bright. Indeed they are brighter
than the reference stars. There are three main types of object:

{\bf Cepheids} distances come from main sequence fits to open clusters.
The VLTI could get $>$1\% distances to $<20$ stars  with parallaxes of
200 -- 1000 \uas and V $<10$. An excellent next step after the
huge ground based studies from microlensing (\cite{be}). 

{\bf RR-Lyrae} distances come from statistical parallax. The VLTI could 
get 1 -- 10\% distances to $<$ 20 stars at $<5$ kpc with V $=8-11$.

{\bf Planetary Nebulae} do not have good distances.
The VLTI could obtain  1 -- 10\% distances for $<40$ with parallax 100 -- 1000
\uas and V $=12-16$.

\subsection{Galaxy Distances from Rotation}

Absolute distances to a few near by Galaxies can be obtained by comparing
Galactic angular and radial velocity rotation rates (van Maanen, 
Reasenberg et al, 1988; \cite{sim}).
Measure both the angular and radial velocities at two places where
the rotation velocity (relative to that galaxies center) 
is expected to be the same, then
solve for the rotation velocity,  inclination and distance.
A transverse velocity of 100 \kms at 1 Mpc is 210 \uas/10 years, which
gives a distance to 5\% per star, and 1\% for 25  stars.
The target A -- F supergiant stars are bright, with M$_v \simeq -8.5$ and
V =21.5 by 10 Mpc, but
few stars in the galaxy will be near to V$<$16 reference stars.

\section{Globular Cluster Distances and Ages}

The next satellite measurements of the cosmic microwave background should
determine the age of the universe to about 1\%.
VLTI could improve globular cluster ages from 10\% today to a few percent.
Improving distances improves the turnoff luminosity, which improves
the mass and thence the age. A 12\% distance error is a 22\% age error
(\cite{ren}).
The VLTI can give 1\% distances at 1 kpc, and 3\% distances with 10 stars
at 10 kpc. Stars with M$_v \simeq -1$ are V $\simeq 14$ at 10 kpc,
providing excellent reference stars. 

Meylan (this meeting) discusses relative proper motions inside clusters.
If we assume globular clusters are isotropic, then a comparison of
radial and angular velocities gives distances.
No global reference frame is needed for this since we are
measuring the dispersion in proper motions, not the absolute values.
Consider a cluster like 47 Tuc with $\sigma_v = 11.5$ \kms, but at 10 times
the distance:  46 kpc. The dispersion of
proper motions will be $\sigma_{\mu} = 53$ \uas/yr. Since stars are close
together in the clusters, we can take advantage of the gains in
relative astrometry with close pairs, perhaps obtaining 2 \uas 
on pairs separated by 1.6 arcsec (V $\simeq 14$). If we then measure
100 stars we get a 10\% error on the $\sigma_{\mu}$.
But gains are slow, since $\sigma_{\mu} \gg $ measurement error, and 
crowding and changing blends may spoil the accuracy.
Astrometric images might be competitive, because they 
capture many stars
at once. A single 8-m should give errors of 150 \uas/yr, 
or $\simeq 3.5 \sigma$ detections of proper motion per star in 10 years, 
which would be difficult for distant clusters,
because the physical dispersion is only 3.5 times
the expected measurement error, which we need to measure accurately and
subtract from the observed dispersion.

\section{Dark Matter in the Galaxy}

The VLTI is well suited to the measurement of the masses and
distances to individual MACHOS in the galactic halo and bulge when they cause
microlensing events.
Microlensing events are seen where
the density of stars is unusually high:
the galactic bulge (detect 50/yr), the Magellanic
clouds (6 -- 8 events in 3 years), and in the near future, the inner
regions of M31.

The bulge events have typical masses of 0.3 M$\sun$, and may be 
caused by ordinary stars in the bar of our Galaxy.
If we knew individual masses and distances we could identify the lens
objects, and later measure masses for planets which we expect will be
found orbiting the lenses (Rhie \& Bennett 1996; Peale 1996).

Events towards the LMC are of unknown origin. Their masses are
around 0.2 M$\sun$, depending on the halo kinematics, and hence
they might be old, cold white dwarfs.

Several possible measurements could be attempted (Miralda-Escud\'e 1996).
The maximum image splitting seen in the sky (Miyamoto \& Yoshii 1996) is
$$2\theta_E = {4 \over c} \sqrt{ GM}\left( {1 \over D_L} - {1 \over D_S} \right)
^{1/2} \leq 2\left({M \over M\sun } {8~kpc \over D_S} \right)^{1/2} ~mas,$$
where M is the lens mass, $D_L$ and $D_S$ are the distances to the lens and 
source star. The event lasts about 40 days (typically 5-100 days, depending on 
the relative velocities and M), and will be hard to observe because the
point spread function is about 1 mas at 1 micron with a 200 m baseline.
 
The centroid shift is 1 -- 2\% of the splitting, or $< 20 - 40$ \uas for a
1 M$\sun$ lens at $D_L=4$ kpc towards a source at $D_S =8$ kpc, which will 
also be hard to measure.

The proper motions of the lens and source are huge: $\simeq 4$ mas/yr, but
measurements will be hard because of the blending of the lens, source and
many other stars in these crowded fields.

Proper motion will separate the lens and source in the sky, to about
18 mas in 5 yr (transverse velocity 100 \kms, $D_L = 6 $ kpc), so that they 
could be resolved. The lens might be faint, V=24 for
$M_v = +10$, and we would like to measure colors to determine a spectroscopic
distance. Hardware and techniques intended to observe planets near to stars
may help with this observation, although it may be built to handle much
larger separations (about 1 arcsec) and higher contrast ratios ($10^9$).

The lens and source parallax are large: 170 \uas at 6 kpc, but again the 
field is crowded, and the lens is very faint.

\section{Structure of Other Galaxies}

The mass to light ratio in dwarf galaxies increases up to 100 as luminosity 
drops. The VLTI could measure the proper motions of many stars to check 
the isotropy of the velocities
which might change M/L determinations by a factor of two (e.g. \cite{eg96}).
The individual stars in a galaxy at 80 kpc (parallax 12 \uas) with internal 
motions of 10 -- 100 \kms would have proper motions of 25 -- 250 \uas.
The distribution of the dispersion in proper motions 
across the galaxy can be used to map out the distribution of dark matter
relative to light, while unusually large motions can indicate mass 
concentrations (black holes).
Alternatively, if we assume the velocities are isotropic, the comparison
of proper motions with radial velocities gives a distance.
Proper motions of stars in large galaxies are hard, but just possible
out to Virgo: 41 \uas in 10 yrs for 300 \kms velocities at 16 Mpc.

\section{Proper Motions of Whole Galaxies}

The proper motions of many whole galaxies are just within the range of
the VLTI.
A 500 \kms velocity at Virgo (16 Mpc) gives 6 \uas/yr, or a 25\% error in 
7 years. The target stars are faint: V=22.5 for $M_v=-8.5$ A -- F Population
I supergiants in spirals and irregulars, and many should be measured per 
galaxy to map out and correct for the internal motions in the galaxies.
A large project could be launched to map the motions of 1000 galaxies, to
reconstruct orbits, identify groups, and map out the mass distribution.
This project requires fringe tracking on very faint stars, since
an 8 kpc galaxy at 16 Mpc covers 1.7 arcmin, and includes on 0.9
foreground stars brighter than V=16 near the Galactic poles, or 16 in the
plane.

\section{Search for Black Holes in Galaxies}

The presence of a massive black hole in a galaxy can be deduced
indirectly from the AGN activity (radio, IR, UV, X-ray and broad emission 
lines), and more directly by sensing the distribution and motions of 
stars and gas (e.g. \cite{miy}).

Images can be used to look for light cusps in the centers of distant
galaxies. If an HST images can reveal a $10^9$ M$\sun$ black hole at 16 Mpc
in images with 100 mas resolution, then the VLTI might see the same at 
400 Mpc with 4 mas images, far enough to sample $10^6$ galaxies.

The VLTI has the astrometric sensitivity to measure the proper 
motions of individual stars near black holes out to Virgo, but this would be
hard because of crowding. Low resolution spectra
of the integrated light from within a few mas of a massive black hole would show
huge velocities, allowing their detection to
great distances (500 \kms at 4 mas for a $2 \times 10^9$ M$\sun$ hole
at 1000 Mpc -- where I scaled distance against angular resolution
from the M87 observations of Ford et al. 1994),
or the detection of lower mass holes in Virgo galaxies
(500 \kms at 16 Mpc for a $3 \times 10^7$ M$\sun$ hole, where $v^2 \propto
M/$[angular resolution]).

\section{AGN and QSOs}

Beyond Virgo, proper motions are too small for the VLTI, and images
and spectra become the main cosmological measurements.
At cosmological distances, 2 mas resolution at 1 micron corresponds to
24 $h^{-1}_{50}$ pc at $z=3$ for $q_0 = 0.5$.
Narrow emission line regions of massive black holes (QSOs) are 100 pc in
size and resolved at all redshifts.
Broad line regions are only 0.1 pc across (2 mas at 10 Mpc)
and hard to resolve, but important because their shape and homogeneity is
unknown. Ward (this volume) discusses IR observations of molecular torii.
Knots in the jets of superluminal sources are probably $< $pc in size, and 
move 0.1 -- 1 mas/yr. Their dynamical range is unknown on mas scales. 
Comparison of optical with radio data relate to the physics of the 
radiation mechanism and the magnetic fields.
If the light from some AGN is dominated by starbursts (Melnick, this
volume), then the star cluster may be resolved.
It would be especially interesting to null out the QSO light to image the
inner structure of high $z$ active galaxies.
High angular resolution images of gravitational lenses provides accurate
astrometry which is the key to the construction of accurate mass models.

\section{Young Galaxies}

Galaxies with cool gas and young stars have a lot of structure;
compare HST images of the near by M100 spiral before and after the
refurbishment. There types of galaxies look fuzzy from the ground 
because of the Earth's atmosphere, and we expect more 
substructure at high redshifts because there is more star formation, and 
lumps from which galaxies are made are not well mixed. Although the high 
galaxies are
extremely faint, and they will have complex shapes, requiring many baselines,
it would be exciting and presumably rewarding
to attempt mas resolution images of even a few.

High redshift galaxies are common, so they can be chosen near to bright
fringe tracking stars. Steidel et al (1996) find 0.4 galaxies at $z \simeq 3$, so
that 3\% of random positions on the sky (e.g near bright stars) have one
such high redshift object.

It would be very interesting to make high resolution images of galaxies
causing absorption lines seen towards bright high redshift QSOs.
Most such QSOs have V$ \geq 18$, but a few at $z \simeq 2$ have V$=16$
for fringe tracking.
We could determine what types of galaxies cause various types of absorption.

\section{Conclusions and Requirements}

Cosmological observations with ground based interferometers are technically
challenging, but the equipment also would benefit most other studies. 
The science is rich and largely unexplored.

There are two main types of measurement: differential
astrometry, and images. The astrometry can give
absolute parallaxes and proper motions when the reference star in the
isoplanatic patch (of the target) has a known parallax and proper motion.
Observations  from space are needed to place these reference stars
in a global (absolute) reference frame.
Astrometry can also measure the dispersion in proper motion of stars in
a globular cluster, dwarf galaxy, or near a massive black hole,
without the need for a reference star in a global reference frame.

Adaptive optics is needed to work in the near-IR, while
laser guide stars are needed on the unit  8-m
telescopes to obtain useful sky coverage. We need 
two perpendicular baselines for astrometry and images, so three unit 
telescopes should be equipped, or only two if the hardware is moved between
telescopes.
 
Every effort should be made to increase throughput and bandwidth so that 
stars fainter than V$=16$ can be used for fringe tracking. Otherwise sky
coverage at the Galactic pole is unacceptably small (3\% with a reference
star within 20 arcsec) and most
individual cosmological targets will not be observable.

Ability to record spectra with a resolution of 100 \kms from light from areas 
of a few mas is important for the
detection of massive black holes and the structure of AGN emission line
regions.

%\newpage

It is a pleasure to thank
Chas Beichman,
Marc Colavita,
Julian Krolik, 
Jordi Miralda-Escud\'e,
Alvio Renzini
especially Mike Shao
for timely information.
%
% ---- Bibliography ----
%

\end{document}